\documentstyle[pra,aps,preprint,tighten]{revtex}
\begin{document}
\draft
\title{Chiral Anomaly and  Spin Gap in One-Dimensional Interacting Fermions}
\author{Naoto Nagaosa and Masaki Oshikawa}
\address{Department of Applied Physics, University of Tokyo,
Bunkyo-ku, Tokyo 113, Japan}
\date{\today}
\maketitle
\begin{abstract}
  Semiclassical approach has been developed for the one-dimensional
interacting fermion systems. Starting from the incommensurate spin
density wave (SDW) mean field state for the repulsive Hubbard model in 1D,
the non-Abelian bosonized Lagrangian describing the spin-charge separation
is obtained.  The Berry phase term is derived from the chiral anomaly, and
we obtain the massless Tomonaga-Luttinger liquid in the single chain case while
the spin gap opens in the double-chain system.
This approach offers a new method to identify the
strong-coupling fixed point, and its relation to the
Abelian bosonization formalism is discussed
on the spin gap state. The generalization to higher dimensions is also
discussed.
\end{abstract}

\narrowtext

\section{Introduction}

  Correlation effects in one-dimension are characterized by the large
quantum fluctuations as well as the strong coupling nature of the
interactions. As for the spin chains, it has been revealed that the
quantal phase, i.e., Berry phase, plays an essential role to determine the
structure of the low energy spectrum \cite{aff1}. Away from half-filling, the
system becomes a Tomonaga-Luttinger liquid (TL-liquid) whose
excitaions are exhausted by the collective modes, i.e., bosonic
degrees of freedom \cite{sol}. This TL-liquid is conveniently described by the
bosonization method where the Hamiltonian is decoupled into
spin and charge parts. All the above features are considered to be
beyond the scope of the conventional mean field + RPA theory which gives
invalid results even qualitatively in 1D. Hence the
interacting 1D Fermion systems are regarded as a laboratory to study the
correlation effects and test many theoretical techniques including the
field theoretical methods and renormalization group (RG) \cite{sol}.

Another aspect of current interest is the spin gap formation and the
possible superconductivity in the double-chain system \cite{dag,ric}.
This system can be regarded as the 1D realization of the short range
resonating valence bond (RVB) scenario proposed for the high-$T_c$
superconductors \cite{and}. For the half-filled case, i.e., without the
charge degrees of freedom, several authors have studied the
double chain systems \cite{str,wat,tsv}, and it is concluded that the spin gap
opens both for antiferromagnetic (AF) and ferromagnetic (F) interchain
exchange couplings. The mechanism for the spin gap is
essentially the same as that proposed by Haldane for $S=1$ AF
Heisenberg chain \cite{hal}.
As for the doped case, mean field type theory \cite{ric},
numerical works \cite{tun,whi}, and analytical studies \cite{fin,fab,khv}
have been done and the persistence of the spin gap
even away from the half-filling has been claimed. It is noted that
the fixed point for the spin gap state is outside the reach of the
perturbative RG, and some assumption about the nature of this strong-coupling
fixed point is nescessary to study its physical properties.

 Lastly the generalization of bosonization and Tomonaga-Luttinger liquid to
higher dimensions has been discussed intensively \cite{bos},
and it is highly
desirable to develope a theoretical framework which is not
restricted to one-dimension but correctly reproduce the results in
one-dimension. It would clarify the role of dimensionality, nesting
condition etc. in the physics of strongly correlated systems.

   In this paper we develop a semiclassical approach to the
interacting 1D fermions. This approach is reliable in higher dimensions,
but has been believed to be a poor approximation even qualitatively in
low dimensionality, especially in one-dimension.
However we will show in this paper that in spite of the large quantum
fluctuations the mean field theory {\it is } a good starting point even in 1D
providing a clear physical interpretation of massless Tomonaga-Luttinger
liquid as well as the strong-coupling fixed point with the spin gap.

For the model with repulsive interactions
the appropriate mean field solution is the spin density wave (SDW)
state \cite{lee}.
We identify the low lying collective modes, i.e., Goldstone modes,
around the mean field solution and derive the effective
Lagrangian for these modes.
By introducing the rotating coordinate system of
the spin \cite{rot,shu}, one can study even the case where the amplitudes
of the collective modes are large and the long range ordering is absent.
In this formulation, the $SU(2)$ connection and hence the gauge
field is naturally derived. This gauge field is interacting with the
1D fermions and the chiral anomaly occurs, which results in the Berry phase
term. The charge part,
on the other hand, is described by the $U(1)$ phase which originates from the
phason degrees of freedom \cite{lee},
and is decoupled with the gauge field mentioned
above. Then the effective Lagrangian is nothing but that of
the non-Abelian bosonization with spin-charge separation \cite{aff2}.
In terms of this analysis, we obtain the massless Tomonaga-Luttinger
liquid for the single-chain case while the spin gap opens for the
double-chain case even away from the half-filling.
This spin gap state is identified in terms of the Abelian bosonization.
By analyzing the correlation functions,
it is found that the low energy dynamics is described in terms of the
bipolaron model \cite{nag1}.
These results in one-dimensional systems are based upon the collective-mode
description of the low energy physics. This is possible only when all the
individual excitations have a gap induced by the order parameter
of the starting long range ordering, e.g.,
spin density wave. In higher dimensions this is not the case in general.
When the nesting condition is well satisfied, however,
the effective action remains similar even in higher dimensions.
When the nesting condition is not satisfied and some parts of the
Fermi surface remain unbroken, the fermionic
field enters into the effective action.
This fermionic field interact with the gauge field, and the
action becomes that of the slave-fermion scheme \cite{shu}.

 The plan of this paper is the followings.
In section II the semiclassical approach with the rotating frame
is developed for one-dimensional systems.
In section III the strong-coupling fixed point with the
spin gap is identified and analyzed in terms of the Abelian bosonization.
A short version of this part has been already published by one of the
present authors (N.N.)\cite{nag1}, and we add the comparison with the
semiclassical approach.
Discussions including the generalization to higher dimensions
and some conclusions are given in section IV.

\section{Semiclassical Approach with Rotating Frame}

We start with the Hubbard model in 1D.
\begin{equation}
H=-t\sum_{i\sigma}(c_{i\sigma}^{\dagger}c_{i+1\sigma+h.c})
-\mu_0\sum_{i\sigma}c_{i\sigma}^{\dagger}c_{i\sigma}+U\sum_in_{i \uparrow}
n_{i \downarrow}
\end{equation}
with the standard notations. The Hubbard interaction is rewritten as
\begin{equation}
Un_{i \uparrow}n_{i \downarrow} ={U \over 2}(n_{i \uparrow}+n_{i \downarrow})
-{U \over 6}(c_{i \alpha}^{\dagger} \vec{\sigma}_{\alpha \beta} c_{i \beta})^2
  \end{equation}
where $\vec{\sigma}_{\alpha \beta}$ is the $(\alpha \beta)$ component of the
Pauli matrix $\vec{\sigma}=(\sigma_x,\sigma_y,\sigma_z)$. Therefore eq.(2)
means that the repulsive interaction favors the formation of local spin moment
$\vec{S_i}={1 \over 2}c_{i \sigma}^{\dagger} \vec{\sigma}_{\alpha \beta}
c_{i \beta}$. By introducing the vector Stratonovich-Hubbard field
$\vec{\varphi}(\vec{r},\tau)$,
the partition function of the system is given by
\begin{equation}
Z=\int Dc^{\dagger}DcD \vec{\varphi}  \exp \biggl[-\int_0^{\beta}Ld \tau
\biggr],
\end{equation}
where
\begin{eqnarray}
L &=&\sum_{i,\sigma}c_{i \sigma}^{\dagger}(\partial_{\tau}-\mu)c_{i \sigma}
-t \sum_{i,\sigma}(c_{i \sigma}^{\dagger}c_{i+1 \sigma}+h.c.)
\nonumber \\
&-& {U \over 3}\sum_{i,\alpha,\beta}\vec{\varphi_i}(\tau)\cdot
c_{i \alpha}^{\dagger}\vec{\sigma}_{\alpha \beta}c_{i \beta}
+{U \over 6}\sum_i\vec{\varphi_i}(\tau)^2
\end{eqnarray}
The mean field state corresponds to the static saddle point for the
$\vec{\varphi}_i(\tau)$ integration, i.e.,
\begin{equation}
{U \over 3}\vec{\varphi}_i^{(s.p.)}
=2m\hat{e}_z \cos(2k_Fx_i) \qquad,
\end{equation}
where $k_F$ is the Fermi wavenumber and the spin polarization is assumed to
be along the $z$ direction, and the amplitude $m$ is determined by the
self-consistency condition. Because the amplitude fluctuation is massive,
we neglect it and consider only the variation of phase $\theta$ and direction
$\vec{n}$ $(|\vec{n}|=1)$ of the order parameter corresponding to the
Goldstone modes of this SDW state \cite{lee}.
\begin{equation}
{U \over 3}\vec{\varphi}_i(\tau)=2m\vec{n}_i(\tau)\cos(\theta(x_i,\tau)
+2k_Fx_i)
\end{equation}
where $\vec{n}_i(\tau)$ and $\theta(x_i,\tau)$ are assumed to be slowly
varying, and we now employ the continuum approximation.
\begin{equation}
L=\int dx
\left[ \begin{array}{c}
R^{\dagger} \\
L^{\dagger}
\end{array} \right] \left[
\begin{array}{cc}
\partial_{\tau} + i e A_0 - i v_F( \partial_x + i e A_x),&
m (\vec n \cdot \vec \sigma ) e^{ i \theta} \\
 m( \vec n \cdot \vec \sigma) e^{ -i \theta},
& \partial_{\tau} +ie A_0 + i v_F (\partial_x + i e A_x)
\end{array} \right]
\left[
\begin{array}{c}
R \\
L
\end{array} \right]
\end{equation}
Here
$
R=\left[
\matrix{
R_{\uparrow} \cr
R_{\downarrow} \cr
}\right]
(L=\left[
\matrix{
L_{\uparrow} \cr
L_{\downarrow} \cr
}\right])
$
represents the right going (left going) branch of Fermion with the linearized
dispersion $\pm v_Fk$ measured from the Fermi wavenumber $\pm k_F$, and the
external electromagnetic field $(A_0,A_x)$ is introduced.
We introduce the polar coordinate $(\zeta,\eta)$ as
$\vec{n}=(\cos \eta \sin \zeta, \sin \eta \sin \zeta, \cos \zeta)$.
Correspondingly we employ the rotating frame whose $z$-axis coincides with
$\vec{n}$ \cite{rot,shu}. Explicitly this rotation can be accomplished by
introducing the $SU(2)$ rotation matrix

\begin{equation}
U \equiv
e^{-i \eta \sigma_z/2}
e^{-i \zeta \sigma_y/2}
e^{i \eta \sigma_z/2}
= \left[
\begin{array}{cc}
z_{\uparrow},&
-z^*_{\downarrow} \\
 z_{\downarrow},
& z^*_{\uparrow}
\end{array} \right]
\end{equation}
where
\begin{equation}
z=\left[
\matrix{
z_{\uparrow}  \cr
z_{\downarrow}  \cr
}\right]
=
\left[
\matrix{
e^{i ( b -\eta)/2} \cos{{\zeta} \over 2}  \cr
e^{i ( b +\eta)/2} \cos{{\zeta} \over 2}  \cr
}\right]
\end{equation}
is the spin 1/2 spinor and
$b$ is corresponding to the gauge choice.
By using this $U$, one can easily obtain
$U \sigma^z U^{\dagger} =\vec{n}\cdot \vec{\sigma}$.
Then the cross term in eq.(7) can be written as
$mR^{\dagger}(\vec{n}\cdot \vec{\sigma})e^{i \theta} L
= m\tilde{R}^{\dagger}\sigma_z \tilde{L}
$ where $\tilde{R} \equiv e^{-i \theta/2} U^{\dagger} R$
and $\tilde{L} \equiv e^{i \theta/2} U^{\dagger} L$
represents the fermions in the rotated frame.
The phason field $\theta(\vec{r},\tau)$ appears as the chiral gauge
transformation. Hence we expect the chiral anomaly.

The chiral anomaly appears as the result of the gauge invariant
regularization of the diverging integrals. One useful way to treat
this anomaly is the
Fujikawa's method \cite{fuj} ( see Appendix).
By using this method and also taking care of the
spin degeneracy factor of 2, the
Jacobian for the change of integral variables is obtained as
\begin{equation}
J=
{{\partial(R,R^{\dagger},L,L^{\dagger})
}\over{
\partial(\tilde{R},\tilde{R}^{\dagger},\tilde{L},\tilde{L}^{\dagger})}}
=\exp[-{{ie} \over {\pi} } \int
dx d\tau\theta(x,\tau)E(x,\tau)]
\end{equation}
where $E(x,\tau)= \partial_{\tau} A_x- \partial_x A_0$ is the electric field.
The above equation describes the acceleration of the phason in terms of
the electric field \cite{sak}. Besides this Jacobian factor,
the Lagrangian in eq.(7)
becomes
\begin{equation}
L=\int dx
\left[ \begin{array}{c}
\tilde{R}^{\dagger} \\
\tilde{L}^{\dagger}
\end{array} \right] \left[
\begin{array}{cc}
\partial_{-} - {i \over 2}\partial_{-}\theta + ieA_{-} +
U\partial_{-}U^{\dagger}, &  m\sigma_z \\
 m\sigma_z, & \partial_{+} + {i \over 2}\partial_{+}\theta
+ ieA_{+}
+ U\partial_{+}U^{\dagger}
\end{array} \right]
\left[
\begin{array}{c}
\tilde{R} \\
\tilde{L}
\end{array} \right]
\end{equation}
where $\partial_{\pm}=\partial_{\tau} \pm iv_F \partial_x$ and $A_{\pm}=A_0 \pm
iv_F A_x$. The explicit expression for $U \partial _{\pm} U^{\dagger}$ is
given as
\begin{eqnarray}
U \partial_{\pm}U^{\dagger} &=&
{i \over 2}(-\partial_{\pm}\eta\sin\zeta\cos b+\partial_{\pm}\zeta\sin b)
\sigma_x
\nonumber \\
&+&{i \over 2}(\partial_{\pm}\eta\sin\zeta\sin b+\partial_{\pm}\zeta\cos b)
\sigma_y
\nonumber \\
&+&{i \over 2}(\partial_{\pm}\eta\cos\zeta+\partial_{\pm}b)\sigma_z \qquad ,
\end{eqnarray}
where the first and second terms represent the spin wave degrees of freedom.
\par\noindent
The coefficient of $\sigma_z$, on the other hand, describes the connection
between the neighboring rotating frame.
Hence we regard it as the $U(1)$ gauge field, i.e.,
\begin{equation}
a_{\pm}=a_0 \pm iv_Fa_x \equiv \partial_{\pm} \eta \cos \zeta +
\partial_{\pm}b.
\end{equation}
The "electric field" $e$ is related to the Skyrmion (instanton) number density
$\vec{n} \cdot (\partial_{\tau} \vec{n} \times \vec{\partial_x}n)$ as
$e = \partial_{\tau} a_x-\partial_x a_0
= \vec{-n}\cdot (\partial_{\tau}
\vec{n} \times \partial_x \vec{n})$.
 From eqs.(11) and (12) there are two massive Dirac fermions coupled with
the internal gauge field $a_{\mu}$, i.e., up spin electron with mass $m$
couples with $ +{1 \over 2}a_{\mu} $ while down spin electron with mass $-m$
couples with $-{1 \over 2}a_{\mu}$.
Hence the effective action $S[a_{\mu}]$
for $a_{\mu}$ is given as
\begin{equation}
e^{-S[a_{\mu}]}=Z(\{a_{\mu}/2 \},m)Z(\{-a_{\mu}/2 \},-m)
\end{equation}
Because we are employing the Euclidean formalism and calculate the partition
function, the complex phase factor is coming from the contour integral
$\int \vec{A}(\vec{r}_i)\cdot d{\vec{r}_i}$ in the first quantization
formalism. Then the relation
$Z(\{-a_{\mu}/2 \},m)=[Z(\{a_{\mu}/2 \},m)]^* $
holds.
Now the chiral anomaly is again relevant. By the chiral gauge transformation
$ \tilde{R} \to e^{{\rm i} \xi/2}\tilde{R}$,
$\tilde{L} \to e^{{\rm -i} \xi/2} \tilde{L}$ with $\xi =\pi$,
the sign of the mass is inverted, i.e., $m \to -m$.
The Jacobian for this transformation is
$J=\exp[- {i \over {4}} \int dx d\tau e(x,\tau)]=\exp[i \pi Q], $
where $Q=\int dxd\tau \vec{n}\cdot (\partial_0 \vec{n} \times  \partial_x
\vec{n})/ 4 \pi$ is an integer called the Skyrmion number.
This relation together with eq.(14) gives
\begin{equation}
e^{-S[a_{\mu}]}=e^{i \pi Q}|Z(\{a_{\mu}/2 \},m)|^2 \qquad ,
\end{equation}
which states that the negative interference occurs
between the topological sectors with even and odd Skyrmion numbers.
This topological term was first derived by
Haldane for the Heisenberg antiferromagnet summing up the Berry
phase terms of the individual spins \cite{hal}.
Our derivation generalizes his results to
(i) itinerant antiferromagnet and (ii) incommensurate case.
We have shown that the chiral anomaly for $A_{\mu}$ describes the acceleration
by the electric field $E$ while that for $a_{\mu}$ the topological term.
One may wonder the validity of the continuum approximation
employed to derive eq.(15). To check this point we have studied the
Berry phase for the adiabatic change of the gauge field $a_{\mu}$ for
several noninteracting electron systems in 1D on a lattice.
We found that,
although the phase of $Z(\{a_{\mu}/2 \},m)$ depends on the
model, the relations
$Z(\{a_{\mu}/2 \}, m) = \exp{(i \pi Q)}Z(\{a_{\mu}/2 \}, -m)$ and eq.(15)
are always valid.
That is, while the Berry phase itself is regularization-dependent,
our calculations leading to eq.(15) is universal.

 We now proceed to complete our derivation of the effective Lagrangian for
$\varphi$ and $\vec{n}$.
For that purpose, we expand the effective action in terms of the
derivative \cite{bra}.
We employ the gauge invariant regularization scheme and
expand with respect to $\partial_{\pm} \theta$
and the spin wave part of $U \partial_ {\pm} U^{\dagger}$ neglecting
the derivative terms of $A_{\pm}$ and $a_{\pm}$.
This procedure can be performed by expanding the $Tr ln$ with
respect to these quantities, and the lowest order term is given by

\begin{equation}
 \sum_{k, \omega_n}
Tr [G_0(k,\omega) V]^2
\end{equation}
where
\begin{equation}
G_0(k,\omega) = { 1 \over { \omega^2 + (v_F k)^2 + m^2} }
\left[
\begin{array}{cc}
i \omega + v_F k, &  m \sigma_z \\
 m \sigma_z, & i \omega - v_F k
\end{array} \right]
\end{equation}
and
\begin{equation}
V =
\left[ \begin{array}{cc}
- {i \over 2} \partial_- \theta + U \partial_-  U^{\dagger}, &
 0 \\
 0, &
- {i \over 2} \partial_+ \theta + U \partial_+ U^{\dagger}
\end{array} \right]
\end{equation}
Becuase $V$ already contains the derivative, we regard $V$ as if it is
a constant matrix to obtain the second order terms with respect to the
derivative. Another comment on the integral in eq.(16) is that
we preserve the Lorentz invariance, i.e., the symmetry
between $\omega$ and $v_F k$.
Then the final result is
\begin{eqnarray}
S &=& \int dx d \tau \biggl[ -i{e \over {2\pi} }
 \theta E+{1 \over {4 \pi v_F} }[(\partial_0 \theta)^2
+(v_F \partial_x \theta)^2] \biggr]
\nonumber \\
&+& i \pi Q +
\int dx d \tau {1 \over { 4 \pi v_F} }[(\partial_0 \vec{n})^2
+(v_F \partial_x \vec{n})^2] ,
\end{eqnarray}
which is the sum of the standard phason Lagrangian plus the nonlinear $\sigma$
model with topological term. We have considered the slowly varying $\theta$
and $\vec{n}$ without assuming the small amplitude of these fluctuations.

  Some remarks are now in order. In our Lagrangian eq.(19) there is
no $U$ appearing, which means that eq.(19) describes the
noninteracting fermions.
This drawback can be easily fixed by taking into account
the forward scattering, i.e., $g_2$ and $g_4$ terms
in the terminology of $g$-ology,  as follows. Note that the density of the
right-going (left-going) fermions $\rho_R$ ($\rho_L$) is given by
$ \rho_R =  \partial_+ \theta /(4 \pi v_F)$
($ \rho_L =  \partial_- \theta /(4 \pi v_F)$).
The forward scattering term is
$U(\rho_R +\rho_L)^2$ which is
translated to
$ U (\partial_+ \theta + \partial_- \theta )^2/(4 \pi v_F)^2$,
which should be added to eq.(19).
Therefore we obtain the action related to $\theta$ as
\begin{equation}
S_{\theta} = \int d \tau dx { 1 \over {4 \pi v_F} }
\biggl[ \biggl( 1 + { U \over { 2 \pi v_F} } \biggr)
(\partial_{\tau} \theta)^2
+ v_F^2 (\partial_{x} \theta)^2 \biggr]
\end{equation}
This action is rewitten in terms of the charge exponent $K_{\rho}$ and
the renormalized charge velocity $v_{\rho}$ as
\begin{equation}
S_{\theta} = \int d \tau dx { 1 \over {4 \pi v_{\rho} K_{\rho} } }
\biggl[ (\partial_{\tau} \theta)^2 + v_{\rho}^2
(\partial_{x} \theta)^2 \biggr]
\end{equation}
where
$K_{\rho} = ( 1 + U/2 \pi v_F )^{-1/2}$
and $v_{\rho} = v_F ( 1 + U/2 \pi v_F )^{-1/2}$.
Thus the Hubbard interaction $U$ manifests itself through
the forward scatterings in the
renormalization of the charge exponent $K_{\rho}$ and
charge velocity $v_{\rho}$.
The spin part, on the other hand, is not affected by $U$.
These features are in consistent with the exact analysis at least
in the small $U$ limit where our approach is expected to be valid.

Now the Lagrangian is actually
equivalent to that of non-Abelian bosonization which respects the
$U(1) \times  SU(2)$ symmetry \cite{aff2}.
Note that the gauge field $a_{\mu}$ representing the Skyrmion number
density is not coupled with the charge phase $\theta$.
The actions for spin and charge are completely decoupled.
In the case of attractive $U$ Hubbard model, the appropriate starting mean
field solution is that of the singlet superconductivity where the
gap is opened. Therefore only the charge degrees of freedom is left
which is described by the Josephson phase which is canonical conjugate to the
phason field described above and the Lagrangian of essentially the same
form as in the repulsive case is obtained. The
charge exponent $K_{\rho}$, however, is larger than unity with the
dominant superconducting fluctuations.

 The periodicity of the lattice has been neglected thus far.
This corresponds to the commensurability pinning of the
phason. At half-filling it is well known that the phason
degrees of freedom is absent because
$Q=2k_F$ is equivalent to $-Q = - 2k_F$.
Therefore the system is insulating,
and only the spin degrees of freedom, i.e.,
$\vec n$ are left as the low lying modes.
This is exactly the Mott insulator in one-dimension.
For other commensurability, e.g., quarter filling case,
the commensurability pinning can be described in the
following way. Let the $Q=2k_F=G/N$ with
$G$ being the reciprocal lattice vector and $N$ an integer.
Because the charge density induced by the SDW is due to the second order
effect and proportional to $\cos(2 Q x + 2\theta)$,
$2MQ = (2M/N)G$ ($M, N$: integers) should be some multiple of $G$ in order that
$2M$-th order energy contributes.
In this case there appears a commensurability energy as
\begin{equation}
\int d x V_{2M} \cos[2 M \theta(x) + \alpha],
\end{equation}
where $V_{2M}$ is some potential energy and $\alpha$ is some phase.
The integer $2M$ is no less than 4, and $V_{2M}$ is irrelevant
when the exponent $K_{\rho}$ is near unity, i.e.,
weak coupling. in which we are interested in this paper.

\par
We now apply our semiclassical method to the double-chain (two-band) model.
 Our interest here is the possible coexistence of the metallic conduction and
the spin gap in the models with repulsive interactions only \cite{ric}.
Now consider the double-chain Hubbard model
coupled with the interchain hopping $t_{\perp}$ and the exchange interaction
$J$.
First consider the case of $t_{\perp} =0$. We can follow the
discussion above and the
saddle point configuration $\vec \varphi^{\rm (s.p.)}$ is
$\vec \varphi_{i}^{(A)} =  \pm \vec \varphi_{i}^{(B)}
\propto m {\hat e_z} \cos(2 k_F x_i)$ depending on the sign of the exchange
$J$.  Here $A$ and $B$ are chain indices.
The effective action for the phason and the spin wave is
the same as eq.(19) for each chain and there appears the
interchain coupling.
\begin{eqnarray}
 -&J& \int d x \vec \varphi^{(A)}(x) \cdot \vec \varphi^{(B)}(x)
\nonumber \\
&\cong& - J \int d x \vec n^{(A)}(x) \cdot \vec n^{(B)}(x)
\cos(\theta^{(A)}(x) - \theta^{(B)}(x) ).
\end{eqnarray}
Because of this coupling the low lying collective modes
also satisfy this relation
$\vec \varphi = \vec \varphi^{(A)} =  \pm \vec \varphi^{(B)}$
, i.e., $\vec n = \vec n^{(A)} =  \pm \vec n^{(B)}$
and $\theta = \theta^{(A)} = \theta^{(B)}$
and the effective action for the phason and the spin wave is
again the same as eq.(19) except the change of the Berry phase term,
i.e., i$\pi Q$ becomes 2i$\pi Q$ ( for $J>0$ ) or 0 ( for $J<0$ ).
In both cases the spin gap opens and the phason mode
gives the metallic conduction. Because of the spin gap,
this state with the spin gap is stable against
at least the small interchain hopping $t_{\perp}$.
Considering that the energy gain due to the spin gap formation
is of the order of $| J_{\perp}|$ while the kinetic energy of the
doped carriers ( concentration $\delta$ ) is of the order of
$\delta t$ or $\delta t_{\perp}$, the spin gap state is expected to be
stable as long as $ \delta \max(t,t_{\perp}) << |J_{\perp}|$.
In the next section we study the various correlation functions in this
spin gap state. For this purpose, the semiclassical method
presented above is not convenient because the canonical conjugate field
$\theta_-$ to the phason field $\theta$ has been already integrated
over when the effective action is derived.
However the considerations above uniquely determine the
strong coupling fixed point which is translated into the language of
Abelian bosonization.
The effect of the interchain hopping $t_{\perp}$ will be again discussed
and it will be shown that it is irrelevant at the strong-coupling fixed point
corresponding to the spin gap state.

\section{ Spin Gap State in Abelian Bosonization}

In this section we relate the spin gap state discussed  in the
previous section to the strong-coupling fixed point in the
Abelian bosonization scheme.
We start with the following Hamiltonian in the Abelian bosonization
formalism \cite{sol}.

\begin{equation}
H=H_A
+H_B
+H_{t_{\perp}}
+H_{J_{\perp}}
\end{equation}
\noindent
where
\begin{equation}
H_i
=v_{\rho} \int dx \biggl[ {1 \over {4\pi\eta_{\rho}} }
\biggl( { {d \theta_+^{(i)} } \over {dx} } \biggr)^2
+\pi \eta_{\rho} P_+^{(i)2} \biggr]
+v_{\sigma} \int dx \biggl[ { 1 \over {4\pi\eta_{\rho} }  }
\biggl( { {d \phi_+^{(i)} } \over {dx} } \biggr)^2
+\pi \eta_{\sigma} M_+^{(i) 2} \biggr],
\end{equation}
\noindent
with $i = A, B$ being the chain index and
\begin{equation}
H_{t_{\perp}}
=-t_{\perp} \int dx \biggl[ \psi^{(A) \dagger}_{R \sigma}(x)
\psi^{(B)}_{R \sigma}(x)
+\psi^{(A) \dagger}_{L \sigma}(x)
\psi^{(B)}_{L \sigma}(x)
+h.c. \biggr],
\end{equation}
\begin{equation}
H_{J_{\perp}}
=-J_{\perp} \int dx \vec{S_{A}}(x) \cdot
\vec{S_{B}}(x).
\end{equation}
\noindent
$\theta_{+}^{(i)}(\phi_{+}^{(i)})$
is the phase variable describing the charge (spin)
degrees of freeedom, and
$P_{+}^{(i)}(M_{+}^{(i)})$
is its canonical conjugate momentum.
$P_{+}^{(i)}(M_{+}^{(i)})$
is related to the phase variable
$\theta_{-}^{(i)} (\phi_{-}^{(i)})$
as
$P_{+}^{(i)}
=- {  { d \theta_{-}^{(i)} } \over  {dx}   }/2\pi$
$( M_{+}^{(i)}
=- { { d \phi_{-}^{(i)} } \over {dx} }/2\pi )$.
In terms of these phase variables, the field operators of the
electrons and the spin are represented
as
\begin{equation}
\psi^{(i)}_{R \sigma}(x)
={1 \over {\sqrt{2 \pi \alpha} }  } \exp \biggl[
i k_{F}x+ {i \over 2 } \{ \theta_{+}^{(i)} + \theta_{-}^{(i)}
+\sigma(\phi_{+}^{(i)} + \phi_{-}^{(i)} ) \} \biggr],
\end{equation}

\begin{equation}
\psi^{(i)}_{L \sigma}(x)
={1 \over {\sqrt{2 \pi \alpha} }  } \exp \biggl[
-i k_{F}x - {i \over 2 } \{ \theta_{+}^{(i)} - \theta_{-}^{(i)}
+\sigma(\phi_{+}^{(i)} - \phi_{-}^{(i)} ) \} \biggr],
\end{equation}

\noindent
and

\begin{equation}
S_+^{(i)}
={1 \over {\pi \alpha} } e^{ - i \phi_-^{(i)} }
[\cos \phi_+^{(i)} + \cos( \theta_+^{(i)}+2k_F x) ],
\end{equation}

\begin{equation}
S_-^{(i)}
={1 \over {\pi \alpha} } e^{ i \phi_-^{(i)} }
[\cos \phi_+^{(i)} + \cos( \theta_+^{(i)}+2k_F x) ],
\end{equation}
\begin{equation}
S_z^{(i)}
={  {\nabla \phi_+^{(i)}}  \over {2\pi} }
+{1 \over {\pi \alpha} }
\cos( \theta_+^{(i)}+ 2k_F x)
\cos \phi_+^{(i)}.
\end{equation}
\noindent
Using these expressions the interchain interactions
$H_{J_{\perp}}$ is written in terms of
$\theta_{\pm}^{(i)}$ and $\phi_{\pm}^{(i)}$ as
\begin{eqnarray}
 H_{J_{\perp}} &=& - {{J_{\perp}} \over { (\pi \alpha)^2} }\int dx
\biggl\{ \cos(\phi_-^{(A)} - \phi_-^{(B)} )
\biggl[ \cos \phi_+^{(A)} \cos \phi_+^{(B)}
 + { 1 \over 2} \cos (\theta_+^{(A)} - \theta_+^{(B)}) \biggr] \biggr\}
\nonumber \\
 &-& {{J_{\perp}} \over 4} \int dx \biggl\{
{ {\nabla \phi_+^{(A)} \nabla \phi+^{(B)} } \over {\pi^2} }
+  { 2 \over {(\pi \alpha)^2} } \cos \phi_+^{(A)}
\cos \phi_+^{(B)} \cos( \theta_+^{(A)} - \theta_+^{(B)} ) \biggr\}
\end{eqnarray}
\noindent
where the rapidly oscillating parts with the wavenumber $\pm 4 k_F$
does not contribute to the integral.

It is obvious from eqs.(6) and (30)-(32) that
the phason field $\theta$ in the previous section is
nothing but $\theta_+$ for each chain.
The director of the spin $\vec n$ is related to the
spin phases $\phi_{\pm}$.
Now we try to identify the spin gap state discussed in the previous section
in terms of Abelian bosonization.
The spin gap state is characterized by the fact that the
relative fluctuations of the spin fields $\vec n$'s and  charge $\theta$'s on
the two chains are massive, i.e., fixed.

Let us first consider the case of half-filling with the
charge degrees of freedom being quenched.
This corresponds to fixing $\theta_+^{(i)}$ in eq.(33),
and the system is described by only the spin phases $\phi$'s.
Several authors have studied this problem \cite{str,wat,tsv}, and
it is concluded the spin gap opens even for infinitestimal $|J_{\perp}|$.
This can be understood as the interchain singlet formation for
antiferromagnetic (AF) $J_{\perp}$ and as the
Haldane gap \cite{hal} for ferromagnetic (F) $J_{\perp}$.
In terms of the Abelian bosonization method these spin gap states are
described as the massive phase of $\phi_+^{(A)} + \phi_+^{(B)}$
and $\phi_-^{(A)} - \phi_-^{(B)}$
for both F and AF $J_{\perp}$ \cite{str,wat}. This is because
$S_+^{(A)} S_-^{(B)} +S_-^{(A)} S_+^{(B)}$
gives rise to $\cos(\phi_-^{(A)} - \phi_-^{(B)})$
and $S_z^{(A)} S_z^{(B)}$
gives rise to $\cos(\phi_+^{(A)} + \phi_+^{(B)})$
and $\cos(\phi_+^{(A)} - \phi_+^{(B)})$.
It is impossible to fix both the canonical conjugate pair
$\phi_-^{(A)} - \phi_-^{(B)}$
and $\phi_+^{(A)} - \phi_+^{(B)}$,
and the spin gap state is realized by fixing
$\phi_-^{(A)} - \phi_-^{(B)}$
and
$\phi_+^{(A)} + \phi_+^{(B)}$.

 From the above construction, the relative direction of $\vec S^{(A)}$ and
$\vec S^{(B)}$ is fixed, i.e.,
$<S^{(A)}_x S^{(B)}_x>$,
$<S^{(A)}_y S^{(B)}_y>$, and
$<S^{(A)}_z S^{(B)}_z>$ are all nonzero, while
there appears no spin moment
( $ < \vec S^{(A)}> = (< \vec S^{(B)} > = \vec 0 $).
Therefore this state is identified as the spin gap state for $\vec n^{(A)}$
and $\vec n^{(B)}$ in the previous section where the relative
phase of them is fixed.

     Now let us turn to the doped case.   The first question in
this case is whether the spin gap survives the doping or not.
When the interchain interactions are treated perturbatively,
the simple power counting arguments give the following conclusions.
As can be seen in eq.(33), the cosine type interactions
in $H_{J_\perp}$
discussed above are multiplied by $\cos(\theta_+^{(A)} - \theta_+^{(B)})$
and their exponents are increased by $2 \eta_{\rho}$.
Therefore the interchain hopping $H_{t_\perp}$ is
more relevant than $H_{J \perp}$, and the conventional
treatment is first to
diagnalize $H_{t \perp}$
to obtain the bonding and anti-bonding bands
and later to take into account the interactions.
However we are interested in the situation where
$J_{\perp}$ is reasonably large, and the electron number is near
the half-filling, i.e., near the Mott insulator.
Let $\delta$ be the concentration measured from half-filling.
Then the characteristic energy due to the phase fluctuation
$ \theta_{\pm}^{(i)} $ is $\max(t, t_{\perp}) \delta$ where $t$ is the
intrachain hopping.
On the other hand, the stabilization energy due to the spin gap
formation is of the order of $|J_{\perp}|$.
Hence if $\max(t, t_{\perp}) \delta \ll |J_{\perp}|$, it is allowed to
consider first the exchange interaction $H_{J_{\perp}}$
and later treat the interchain hopping $H_{t_{\perp}}$
as a perturbation.

 In the case of $t_{\perp} =0$, the effect of the doping is
summarized in the factor
$\cos(\theta_+^{(A)} - \theta_+^{(B)})$ which appears in eq.(33).
If this factor gives the finite expectation value, i.e., the
combination $\theta_+^{(A)} - \theta_+^{(B)}$ is fixed and massive,
the dynamics of the spin phases $\phi$'s remains essentially the same
and the spin gap persists.
On the other hand, if the spin gap persists, i.e.,
$\phi_-^{(A)} - \phi_-^{(B)}$ and $\phi_+^{(A)} + \phi_+^{(B)}$
remain fixed, $\cos(\theta_+^{(A)} - \theta_+^{(B)})$ alone is the
relevant perturbation and $\theta_+^{(A)} - \theta_+^{(B)}$
is fixed, which means that the spin gap state is self-consistent.
This state is identical with the spin-gap fixed point found by
Khveshchenko and Rice \cite{khv}.
This is also identified with the doped spin gap state in section III
where $\theta^{(A)}-\theta^{(B)}$ is fixed and
$<\vec n^{(A)} \cdot \vec n^{(B)}>$ is nonzero.
 Because $\nabla \theta_+^{(A)}$
($\nabla \theta_+^{(B)}$) is the slowly varing part of the
charge density on the chain $A (B)$ \cite{sol}, this  means that the
charge balance between the two chains is kept.
If the carriers are doped with different concentration,
$\theta_+^{(A)} - \theta_+^{(B)}$ is proportional to the
cordinate $x$ along the chain and
$\cos(\theta_+^{(A)} - \theta_+^{(B)})$ does not contribute to the
integral. The spin gap disappears immediately and it is expected that
the system is described as the massless
Tomonaga-Luttinger liquid in this case \cite{ued,kawa}.

 This spin gap state has stability against small
interchain hopping $t_{\perp}$ because $H_{t_{\perp}}$ is the irrelevant
perturbation in this case, i.e., its expectation value
vanishes and correlation function decays exponentially.
Also it should be noted that the two chains need not be exactly equivalent.
If the chemical potential difference $\Delta \mu$ between the two chains
are small enough compared with $|J_{\perp}|$, the stabilization
energy of the gap formation prefers the equal concentration of the
holes on each chain.

     Now we consider the physical properties of the doped spin gap state.
As has been discussed above, this state is
characterized as the fixed and massive state of the combinations
$\phi_+^{(A)} + \phi_+^{(B)}$, $\phi_-^{(A)} - \phi_-^{(B)}$,
and $\theta_+^{(A)} - \theta_+^{(B)}$.
In this state the canonical conjugate fields to those fixed ones,
i.e.,
$\theta_+^{(A)} - \theta_+^{(B)}$,
$\theta_-^{(A)} + \theta_-^{(B)}$,
and $\theta_-^{(A)} - \theta_-^{(B)}$,
 are uncertain and their exponentiated operators have zero
expectation value and exponentially decaying correlation functions.
The only massless mode is then $\theta_{\pm}^{(A)} + \theta_{\pm}^{(B)}$,
i.e., the in-phase charge fluctuations.
 From these considerations the following conclusions are derived
immediately \cite{nag1}.

\noindent
(1) Among numerous order parameters constructed as the product of the
two fermion fields
, only the interchain singlet superconductivity $O^{AB}_{SS}$
given below shows the power-law behavior while all the others show exponential
decay.
\begin{equation}
O_{SS}^{AB}
= \psi_{R \sigma}^{(A)}
 \psi_{L - \sigma}^{(B)}
= { 1 \over {2 \pi \alpha} } \exp { i \over 2} [
(\theta_+^{(A)} - \theta_+^{(B)}) + (\theta_-^{(A)} + \theta_-^{(B)})
+ \sigma (\phi_+^{(A)} + \phi_+^{(B)})
+ \sigma (\phi_-^{(A)} - \phi_-^{(B)}) ].
\end{equation}

\noindent
(2) When one goes further to the product of four fermion fields,
the $4k_F$ charge density wave $O^{ABAB}_{\sigma \sigma'}$
given below shows the power-law behavior.
\begin{eqnarray}
O^{ABAB}_{\sigma \sigma'} &=& \psi^{(A) \dagger}_{R \sigma}
\psi^{(B) \dagger}_{R \sigma'} \psi^{(A)}_{L \sigma'}
\psi^{(B)}_{L \sigma}
\nonumber \\
&=& \biggl( { 1 \over {2 \pi \alpha} } \biggr)^2
 \exp \biggl[ - 4 i k_F x - i( \theta_+^{(A)}+ \theta_+^{(B)})
- i \sigma \delta_{\sigma \sigma'}( \phi_+^{(A)}+ \phi_+^{(B)})
- i \sigma \delta_{\sigma - \sigma'}( \phi_-^{(A)}- \phi_-^{(B)})
\biggr].
\end{eqnarray}

\noindent
(3) The exponents $K_{SC}$ and $K_{CDW}$ for the two order parameters
defined in eqs.(34) and (35) respectively are determined by the only
massless field $\theta_{\pm}^{(A)} + \theta_{\pm}^{(B)}$, and satisfies the
duality relation $K_{SC} \cdot K_{CDW} = 1$ \cite{com}. And it is noted
that the wavenumber $4 k_F$ is corresponding to the
$" 2 k_F "$ of the spinless hard core boson.
Note that the $2 k_F$ charge density wave of the fermion does not show
power-law correlation.
Therefore this spin gap state in the double chain system is distinct from the
Luther-Emery type state in the single-chain state \cite{oga,ima}.
The Luther-Emery state for negative $U$ Hubbard model is understood as
the fluctuating singlet superconducting state as has been discussed in
sention II.
These two facts means that the low energy dynamics of the spin gap state is
described as the bipolaron along the lung of the ladder. This is also
consistent with the above conclusion that the spin gap persists only when the
two chains are doped equally.
Recent numerical study of the double-chain Hubbard model \cite{whi} suggests
that $K_{SC}$ is 2 which means $K_{CDW} = 1/2$, i.e., the CDW is
dominating.
However we have more chance to obtain superconducting state in the $t$-$J$
ladder
model because the naive expectation is that the product of two fermion
operators have small scaling dimensions than that of four fermion operators.
In this picture the charge carriers in the
spin gap state is the charge $2e$ bipolaron.

\section{Discussion}

In this section we discuss the implications of the above results for
higher dimensions. The Stratonovich-Hubbard transformation
( eqs. (1)-(4) ) and the rotating frame method ( eqs.(8)-(9) )
can be used in any dimensions.
In eq.(4) there occurs a competition between the kinetic energy
$t$ and the local magnetic field $ 2U \vec \varphi_i/3$. The former
prefers the nonmagnetic state due to the Fermi degeneracy, while the
latter induces the magnetic moments. The criterion for the
weak and strong correlation is the relative magnitudes of these
two, i.e., the bandwidth $W \sim zt$ ($z$: number of the nearest neighbor
sites) and the average strength of the local magnetic field
$ F = 2 U <|\vec \varphi_i | >/3$.

When $F<<W$ the laboratory frame with the fixed spin axis is the
appropriate coordinate in which the kinetic energy Hamiltonan is
easily diagonalized. The local magnetic field is a small perturbation
which slightly induces the magnetic moments. This can be treated as a
weak scattering of the fermions due to the local magnetic field.
The density of states remains essentially the same as the noninteracting case.

When $F>>W$, on the other hand, the fermion spin at site $i$ is forced to be
parallel to the direction of the local magnetic field
$\vec n_i = \vec \varphi_i/|\vec \varphi_i|$.
Therefore one must take into account the local magnetic field first by
employing the rotating frame whose $S^z$-axis coincides with
$\vec n_i=( \cos \eta \sin \zeta, \sin \eta \sin \zeta, \cos \zeta)$.
This can be done similarly to eqs.(8) and (9) by introducing the
$2 \times 2$ unitary matrix $U_i$.
\begin{equation}
U_i \equiv
e^{-i \eta_i \sigma_z/2}
e^{-i \zeta_i \sigma_y/2}
e^{i b_i \sigma_z/2}
= \left[
\begin{array}{cc}
z_{i \uparrow},&
-z^*_{i \downarrow} \\
 z_{i \downarrow},
& z^*_{i \uparrow}
\end{array} \right]
\end{equation}
where
\begin{equation}
z_i=\left[
\matrix{
z_{i \uparrow}  \cr
z_{i \downarrow}  \cr
}\right]
=
\left[
\matrix{
e^{i ( b_i -\eta_i)/2} \cos{{\zeta_i} \over 2}  \cr
e^{i ( b_i +\eta_i)/2} \cos{{\zeta_i} \over 2}  \cr
} \right].
\end{equation}
Then we define the fermion field in the rotated frame as

\begin{equation}
 c_i =
\left[
\matrix{
c_{i \uparrow}  \cr
c_{i \downarrow}  \cr
}\right]
= U_i \bar{c} = U_i
\left[
\matrix{
\bar{c}_{i \uparrow}  \cr
\bar{c}_{i \downarrow}  \cr
}\right].
\end{equation}
The local "magnetic field" is always along the $+z$
direction for the new fermion $\bar{c}$.
In this case the density of states for the fermions $\bar{c}^{\dagger}$
and $\bar{c}$ is split into two according to their spins
with the gap of the order of $F$.
This is nothing but the upper and lower Hubbard bands with the bandwidth
of the order of $W$. Note that we do not assume the magnetic ordering.
For each Hubbard band, the number of available states is that of the
lattice sites $N$.
Then for the half-filled case, only the up-spin band, i.e.,
lower Hubbard band, is occupied, and the system becomes the Mott insulator.
When the holes are doped, there appears the small hole pocket near
the top of the lower Hubbard band.
In this case the upper Hubbard band is irrelevant for the low energy
behavior of the system, and we neglect it for the moment.
Then the system is described by the spinless fermion
$f_i = \bar{c}_{i \uparrow}^{\dagger}$ and its conjugate.
The original electron field $c_i$ is given from eq.(38) as
\begin{equation}
c_{i \sigma} = z_{i \sigma} f_i^{\dagger},
\end{equation}
which is exactly the slave-fermion decomposition \cite{shu}.
The low energy dynamics is described in terms of the spinless fermion
for charge degrees of freedom and the spin field $\vec n$ (or $z_{\sigma}$).
In this case the gauge field is coupled with the spinless fermion.

The above discussion did not include the detailed band structure and the
shape of the Fermi surface in the momentum space.
When the nesting condition is satisfied, it is possible that the
gap opens up even when $F<<W$ if one chose the spatial pattern of
$\vec \varphi_i$ appropriately.
The one-dimensional system discussed above is the typical example of
this case, where the "Fermi surface" is the two points
with perfect nesting. We concentrate on the Fourier components
of $\vec \varphi_i$ near $\pm Q = \pm 2 k_F$, and the
low lying excitations are exhausted by the collective modes.
These collective modes are nothing but the variables describing the
Tomonaga-Luttinger liquid, and the spin-charge separation occurs when the
forward scatterings are taken into account.
Therefore it represents the effects of the correlation on the
collective coordinates.
When the nesting condition is well satisfied, this scenario remains basically
the same even in higher dimensions.
Of course the long range ordering generally occurs in higher diemnsions and
the collective modes can be treated in terms of the standard random phase
approximation (RPA) when the fluctuations are small.
However, near or above the transtion temperature, or in the case where
the long range ordering is suppressed by some reason, e.g., frustration
in the interactions, the fluctuations are large and one had better
describe the system in the rotating frame similarly to the case of
strong correlation.
Therefore the system with the nested Fermi surface
share some features with the strongly correlated system when the
long range ordering is absent, and will contribute to the understanding
of the role of the collective modes in the physics of the strongly
correlated systems.

\acknowledgements

The authors acknowledge T.M.Rice, H.Tsunetsuge, K.Ueda, M.Imada,
Y.Kitaoka, M.Takano, E.Daggoto, X.G.Wen T.K.Ng for useful discussions.
This work is supported by Grant-in-Aid for
Scientific Research No. 06243103
from the Ministry of Education, Science, and Culture of Japan.
\vfill\eject
\appendix
\section{Fujikawa's method for Chiral Anomaly}
In this appendix we give a brief description of Fujikawa's
method \cite{fuj} applied to the Dirac fermions in (1+1)-dimensions.
Here we take the unit $v_F = 1$ and let $r=(\tau, x)$ be
the two-dimensional coordinate.
$\psi(r)=\left[
\matrix{
R(r)  \cr
L(r)  \cr
}\right] $
is the two-component spinor, and
$ \bar{\psi}(r) = [ L^{\dagger}(r), R^{\dagger}(r)]$.
We take the chiral representation where $\gamma_5 = \sigma_z$.
Then the chiral gauge transformation is
given by
\begin{eqnarray}
\psi(r) &\to& \psi'(r) = e^{ i \alpha(r) \gamma_5} \psi(r)
\nonumber \\
\bar{\psi}(r) &\to& \bar{\psi}'(r) = \bar{\psi}(r) e^{ i \alpha(r) \gamma_5}
\end{eqnarray}
We are interested in the Jacobian corresponding to the
change of integration variables, i.e., chiral gauge
transformation given above.
For this purpose let us expand $\psi(r)$ in terms of the eigenfunction
$\varphi_m(r)$ of the Dirac operator
$\hat{D} =  \left[
\begin{array}{cc}
0, &  D_+ \\
 D_-, & 0
\end{array} \right]$
 with $D_{\pm} = \partial_{\pm} + iA_\pm $.
 \begin{equation}
\hat{D} \varphi_m(r) = \lambda_m \varphi_m(r)
\end{equation}
and
\begin{equation}
\psi(r) = \sum_{m} a_m \varphi_m(r).
\end{equation}
The the functional integral ${\it D} \psi(r)$ can be
replaced by
$\Pi d a_m$.
Then also the transformed $\psi'$ can be expanded in terms of $\varphi_m$ as
\begin{equation}
\psi'(r) = \sum_{m} a'_m \varphi_m(r).
\end{equation}
with $a'_m$ being given by
\begin{equation}
a'_m = \sum_n \int d^2 r \varphi^{\dagger}_m(r)
e^{i \alpha(r) \gamma_5} \varphi_n(r)
\cdot a_n \equiv \sum_n C_{mn} a_n.
\end{equation}
Coresponding to this change of integral variables
\begin{equation}
\Pi_m d a'_m = [{\rm det} C_{mn}]^{-1} \Pi_n d a_n.
\end{equation}
Assuming small $\alpha(r)$,
the determinant is explicitly given by
\begin{eqnarray}
[{\rm det} C_{mn} ]^{-1} &=&
{\rm det} \biggl[ \delta_{mn} +
i \int d^2 r \alpha(r) \varphi^{\dagger}_m(r)
\gamma_5  \varphi_n(r) \biggr]
\nonumber \\
&=& \exp \biggl[
-i \int d^2 r \alpha(r) \sum_n \varphi^{\dagger}_n(r)
\gamma_5  \varphi_n(r) \biggr]
\nonumber \\
&=& \exp \biggl[
-i \int d^2 r \alpha(r) G(r) \biggr].
\end{eqnarray}
The function $G(r)$ is calculated as
\begin{eqnarray}
G(r) &=&
\sum_n \varphi^{\dagger}_n(r) \gamma_5 \varphi_n(r)
\nonumber \\
&=& \lim_{M \to \infty}
\sum_n \varphi^{\dagger}_n(r) \gamma_5 e^{-(\hat{D}/M)^2} \varphi_n(r)
\nonumber \\
&=& \lim_{M \to \infty}
Tr \int { {d^2 k} \over {(2 \pi)^2} }
\gamma_5 e^{-i k r} e^{-(\hat{D}/M)^2} e^{i k r},
\end{eqnarray}
where we have introduced the gauge invariant convergence factor
$e^{-(\hat{D}/M)^2}$
to remove the ultraviolet divergence.
Now we examine $\hat{D}^2$.
It is easy to obtain
\begin{equation}
\hat{D}^2 = ( \partial_\tau +i A_0 )^2 +
( \partial_x +i A_x )^2 + \gamma_5 F_{0 1}
\end{equation}
where $F_{01} = \partial_\tau A_x - \partial_x A_{\tau}$
is the electric field.
Then the function $G(r)$ can be calculated as

\begin{eqnarray}
G(r) &=& \lim_{M \to \infty}
Tr \int { {d^2 k} \over {(2 \pi)^2} }
\gamma_5 \exp \biggl[ -\sum_{\mu = \tau,x} (k_\mu + A_{\mu} )^2/M^2
- F_{01}\gamma_5/M^2 \biggr]
\nonumber \\
&=& \lim_{ M \to \infty}
- {{2 F_{01} } \over {M^2}} \int { {d^2 k} \over {(2 \pi)^2} }
\exp \biggl[ -\sum_{\mu = \tau,x} (k_\mu + A_{\mu} )^2/M^2 \biggr]
\nonumber \\
&=& - {{F_{01}} \over {2 \pi}}
\end{eqnarray}
Putting  eq.(A10) into eq.(A7) we obtain
\begin{equation}
[{\rm det} C_{mn} ]^{-1} =
\exp \biggl[ i \int d^2 r \alpha(r) {{ F_{01}(r)} \over {2 \pi}} \biggr].
\end{equation}
Similar calculation can be done also for ${\it D} \bar{\psi}$
and the same factor as eq.(A11) is obtained.
Then the Jacobian $J$ for the chiral gauge transformation
is given by
\begin{equation}
J =
\exp \biggl[ i \int d^2 r \alpha(r) {{ F_{01}(r)} \over {\pi}} \biggr],
\end{equation}
which is Fujikawa's Jacobian.


\begin{references}

\bibitem{aff1}For a review see I.Affleck, J. Phys. Condens. Matter
{\bf{1}}, 3047 (1989).

\bibitem{sol} J.Solyom, Adv. Phys. {\bf{28}}, 209 (1979);
V.J.Emery, in {\it Highly Conducting One-Dimensional Solids},
edited by J.T.Devreese et al. (Plenum, 1979);
H.Fukuyama and H.Takayama, in {\it Dynamical Properties of
Quasi-One-Dimensional Conductors}, edited by P.Monceau ( Reidel, 1984).

\bibitem{dag}E.Dagotto, J.Riera, and D.J.Scalapino, Phys. Rev. B{\bf{45}},
5744 (1992); T.Barns, E.Dagotto, J.Riera, and E.Swanson,
Phys. Rev. B{\bf{47}}, 3169 (1993).

\bibitem{ric}T.M.Rice, S.Gopalan, and M.Sigrist, Europhys. Lett. {\bf{23}},
445(1993); T.M.Rice et al, in {\it Correlation Efects in low-Dimensional
Electron Systems}, edited by A.Okiji and N.Kawakami (Springer-Verlag, 1994)
p177.

\bibitem{and}P.W.Anderson, Science {\bf{235}}, 1196 (1987).

\bibitem{str}S.P.Strong and A.J.Millis,
Phys. Rev. Lett.{\bf{69}}, 2419 (1992).

\bibitem{wat}H.Watanabe, K.Nomura, and S.Takada, J. Phys. Soc. Jpn. {\bf{62}},
2845 (1993).

\bibitem{tsv}A.M.Tsvelik, Phys. Rev. Lett. {\bf{72}}, 1048 (1994).

\bibitem{hal}F.D.M.Haldane, Phys. Lett. {\bf{93A}}, 464 (1983);
Phys. Rev. Lett. {\bf{50}}, 1153 (1983).

\bibitem{tun}H.Tsunetsugu, M.Troyer, and T.M.Rice, Phys. Rev. B{\bf{49}},
16078 (1994).

\bibitem{whi}R.M.Noack, S.R.White, and D.J.Scalapino, Phys. Rev. Lett.
{\bf{73}}, 882 (1994) and
unpublished.

\bibitem{fin}A.M.Finkel'stein and A.I.Larkin, Phys. Rev. B{\bf{47}},
10461 (1993).

\bibitem{fab}M.Fabrizio, A.Palora, and E.Tosatti, Phys. Rev. B{\bf{46}},
3159 (1992); M.Fabrizio, Phys. Rev. B{\bf{48}}, 15838 (1993).

\bibitem{khv}D.V.Khveshchenko and T.M.Rice,  Phys. Rev. B{\bf{50}}, 252 (1994);
D.V.Khveshchenko,  Phys. Rev. B{\bf{50}}, 380 (1994).

\bibitem{bos}A.Luther, Phys. Rev. B{\bf{19}}, 320 (1979);
A.Houghton and B.Marston, Phys. Rev. B{\bf{48}}, 7790 (1993);
A.H.Castron Neto and E.H.Fradkin, Phys. Rev. Lett. {\bf{72}}, 1393 (1994)..

\bibitem{lee}P.A.Lee, T.M.Rice, and P.W.Anderson, Solid State Commun.
{\bf{14}}, 703 (1974).

\bibitem{rot}V.Korenman, J.L.Murray, and R.E.Prange, Phys. Rev.
B{\bf{16}}, 4032 (1977).

\bibitem{shu}H.J.Shulz, Phys. Rev. Lett.{\bf{65}}, 2462 (1990).

\bibitem{aff2}I.Affleck, in {\it Fields, Strings and Critical Phenomena},
edited by E.Brezin and J.Zinn-Justin (Elsevier, Amsterdam, 1989), p.563.

\bibitem{nag1}N.Nagaosa,  Solid State Commun.
{\bf{94}}, 495 (1995).

\bibitem{fuj}K.Fujikawa, Phys. Rev. D{\bf{21}}, 2848 (1980).

\bibitem{sak}B.Sakita and K.Shizuya, Phys. Rev. B{\bf{42}}, 5586 (1990).

\bibitem{bra}S.A.Brazovskii and I.E.Dzyaloshinskii,
Sov. Phys. JETP {\bf{44}}, 1233 (1976).

\bibitem{ued}K.Ueda, in {\it Correlation Efects in low-Dimensional
Electron Systems}, edited by A.Okiji and N.Kawakami (Springer-Verlag, 1994)
p107.

\bibitem{kawa} S.Fujimoto and N.Kawakami, to appear in J.Phys. Soc. Jpn..

\bibitem{com}The author acknowledges H.Tsunetsugu for suggesting this point.

\bibitem{oga}M.Ogata, U.M.Luchini, and T.M.Rice,
Phys. Rev. B{\bf{44}},
12083 (1991).

\bibitem{ima}M.Imada, Phys. Rev. B{\bf{48}}, 550 (1993).
\end{references}
\end{document}